\documentclass[11pt]{article}
\pdfinfoomitdate=1 \pdfsuppressptexinfo=-1 \pdftrailerid{}
\usepackage[letterpaper,margin=1in]{geometry}
\usepackage[T1]{fontenc}
\usepackage[utf8]{inputenc}
\usepackage{mathptmx}

\usepackage{amsmath}
\usepackage{graphicx,booktabs,array}
\usepackage[authoryear,round]{natbib}
\usepackage{setspace}
\usepackage{enumitem}
\usepackage[hidelinks]{hyperref}
\usepackage{url}
\usepackage{caption}
\usepackage{float}
\usepackage{pdfpages}
\setlist{itemsep=2pt,topsep=4pt}

\hypersetup{pdftitle={Who Fills the Gap? Role Similarity and Teammates' Minutes When NBA Players Are Ruled Out},pdfauthor={Tianyu Guan and Yang Zhou}}
\title{Who Fills the Gap? Role Similarity and Teammates' Minutes When NBA Players Are Ruled Out}
\author{Tianyu Guan$^{1,2,*}$ \quad Yang Zhou$^{2}$\\[6pt]
{\small $^{1}$Department of Mathematics and Statistics, York University, Toronto, Canada}\\
{\small $^{2}$TY Sports Analytics Inc., Canada}\\
{\small $^{*}$Corresponding author: \href{mailto:tguan@yorku.ca}{tguan@yorku.ca}}}
\date{}
\begin{document}
\maketitle
\doublespacing
\begin{abstract}
\noindent When a core player is ruled out before an NBA game, do teammates whose statistical role resembles that of the removed set (every rotation player ruled out, about three per case) play more? Using injury reports stamped at least six hours before tip-off and rosters reconstructed after the season, we study 2,441 team-games from 2023--24 to 2025--26 in which a core player was ruled out for a medical reason. Among teammates in the same game, one within-case standard deviation more similarity in role profile (three-point attempts, assists and rebounds per minute) goes with about 0.63 more minutes, holding reference minutes and headroom fixed ($p < 0.0001$); this compares teammates within a game, not the minutes gained because of the absence. The pattern holds in two earlier seasons and in exploratory within-player and without-and-with comparisons. Whether similarity adds to listed position is not settled. The association is small and yielded no statistically detectable reduction in the error of the retrospective minute forecasts we tested. Role coverage, how much of the missing minutes and roles the remaining rotation could restore, showed no consistent incremental association with team role output or the final margin.

\end{abstract}
\clearpage

\section{Introduction}

From 2023--24 to 2025--26, the league's official pregame injury report ruled out at least one of a team's three core players (those with the most reference minutes; Section~3.1) for a medical reason in more than a third of team-games (2,441 of the 6,722 with a report filed by our cutoff). The coach then decides who plays the missing minutes. This paper asks whether the allocation follows statistical role: whether teammates whose profile resembles the absent players' play more. The absent players are the removed set, every rotation player ruled out for any reason (3.0 per case on average; Section~3.3), not only the core player who makes the team-game a case, and similarity is to their combined profile. Which teammate replaces the core player himself is a different question, which this design does not answer; the pair comparison of Section~5.2 comes closest.

Absences matter. Teams that lose more games to injury win less often \citep{Podlog2015Injuries}, the in-game exit of a player who is both captain and All-Star lowers teammates' plus-minus \citep{Kocsoy2024Captains}, and betting prices move when injury news arrives \citep{Sauer2003PlayerInjuries,Dare2015PlayerAbsence}. Teammates adjust. In the NHL, the absence of a substitutable co-worker raises established teammates' working time \citep{HoeyPeetersVanOurs2023}. In baseball, same-position replacements gain opportunity after star injuries \citep{Nishihata2026ValueChanceV2}, and in the NBA, after in-game injuries of high performers teammates keep their division of tasks but produce less, and relatively skilled players step in \citep{WegelinOrlowskiDietl2022}. These designs group teammates by task, by listed position or by overall quality, and data-driven classifications have already moved beyond the five positions \citep{Muniz2022WeightedNetwork}. What has not been measured is, first, whether in games with a known pregame absence a teammate plays more the more closely his statistical profile resembles that of the absent players and, second, whether that carries information beyond listed position and past substitution.

For every team-game of those three seasons we take the last official injury report whose header time is at least six hours before tip-off, each player's recent games, and the roster, which we reconstruct after the season where the league's records are incomplete (Section~2.2). We call a team-game a case when that report rules out, for a medical reason, at least one of the team's three core players. Reference minutes divide the game's 240 minutes among the roster according to each player's recent games (Section~3.1). Two earlier seasons (2021--22 and 2022--23), in which our role measures had never been related to any outcome, serve as a second sample. We planned three tests.

\begin{itemize}
  \item Allocation. In cases, do teammates whose statistical role is closer to that of the absent players play more minutes relative to their reference minutes?
  \item Role output. When the retained roster can cover less of a missing role, does the team produce less of that role in the game?
  \item Margin. Does the roster's ability to cover the missing roles explain the final margin, beyond the value of the players who are missing?
\end{itemize}

The first test uses role similarity, the cosine between a teammate's standardised per-minute rates of three-point attempts, assists and rebounds and those of the absent players (Section~3.3). The other two use role coverage, the share the retained rotation could restore: of each role for the role-output test ($c_k$), and jointly, of the absent players' minutes and every role at once under one budget of extra minutes, for the margin test; no player may exceed his 90th percentile of minutes (at most 40).

The main finding is that minutes tilt toward similar teammates: within a case, a teammate one within-case standard deviation more similar to the removed set plays about 0.63 more minutes than a teammate with the same reference minutes and headroom. This is a difference between teammates of one game, not the minutes a teammate gains because of the absence, which only the exploratory comparisons of a teammate's games without and with a given core player approach, and less precisely (Section~5.2). The association recurs in two earlier seasons; it is small and did not yield a statistically detectable reduction in the error of the retrospective forecasts we tested. Whether it carries information beyond listed position is left open: part of it survives position controls over all cases, but in cases whose absent players share one listed position and in pairs with both players' effects the estimates fall to 0.20 and $-$0.02 minutes per unit and are imprecise. The second finding is negative: the planned role-output test misses its threshold ($p=$ 0.070), and role coverage adds no stable information about team role output beyond the absent players' share of each role, or about the final margin beyond the value of the missing players (Section~9.3). Both rest on a point-in-time reconstruction of the injury reports that this paper shares with a companion paper (Section~2).

\subsection{Related work}

\textbf{Absences, teammates and substitutes.} An injured second-tier star lowers a European football team's goal difference \citep{Jedelhauser2023Overshadowed}, and on NBA teams with a single star, losing a star who is a generalist across play types costs more than losing a specialist \citep{Nalbantis2026Generalist}. Who replaces whom has been studied mostly by position: when COVID-19 kept foreign players out of the Chinese Basketball Association, domestic players at the affected positions gained minutes \citep{Jin2026PlayingTime}. Outside sport, coworkers in the same occupation earn more after a worker dies, evidence of substitution between workers and of its cost \citep{JagerHeining2022Substitutable}, and in Danish small firms parental leave costs firms little except where no coworker in the same occupation can cover \citep{Brenoe2024ParentalLeave}; both point to the availability of close substitutes, which our coverage measure quantifies for a roster. Absences also change how teams work together \citep{ChenGarg2018Dancing,Stuart2017StructuralDisruption}, and give rookies playing time that affects their later contracts \citep{Hoey2026Exposure}. A recent NBA study, \citet{Krishna2026Workload}, uses the number of a player's absent teammates in the games of 2017--18 to 2022--23 as an instrument for his minutes, finds that workload raises injury risk and that coaches rest players after shocks to that risk; it counts absent teammates without regard to who they are and does not ask which teammates absorb the minutes. Here the absence is fixed by the pregame report, and the question is how the freed minutes are divided among the retained players according to their role similarity to the absent ones. Absences are not random: whether NBA stars miss a game depends on injuries, back-to-backs, the point in the season and the opponent \citep{Reilly2023StarsOut}. Betting markets react to injury and illness news, though not always at once \citep{FischerSchmal2025Pricing}.

\textbf{Roles beyond listed position.} Data-driven roles replace the five positions in many studies: clustering yields thirteen \citep{Alagappan2012Positions}, self-organising maps with fuzzy clustering five \citep{Bianchi2017Role}, weighted networks eight \citep{Muniz2022WeightedNetwork}, and principal components with $k$-means five in 2025--26 \citep{RodriguezIdiazabal2026Roles}; cosine distance has been used to cluster lineup-based playing styles \citep{Guo2025Versatility}. Listed position nonetheless stays informative: it still accounts for about a quarter of the variation in players' statistical profiles, although dispersion within positions has grown \citep{Durmus2026Positions}, and archetypes add only modest gains over positions in models of team defence \citep{South2024Paradox}. Similarity on components of tracking statistics has been proposed to find cheaper replacements \citep{Bruce2016Scalable}. A critical integrative review of NBA analytics proposes a framework for judging whether evidence is ready to inform decisions and discusses role substitution and player availability \citep{Zhou2026Decisions}. We ask instead whether, after a pregame absence, teammates' minutes are associated with their continuous role similarity to the absent players, and whether that association goes beyond listed position.

\textbf{Allocation of playing time.} Whether draft position affects minutes beyond performance is disputed \citep{Staw1995SunkCosts,Camerer1999Escalation}; adjusting for the games a player was available and using a discontinuity at draft-position thresholds, \citet{Leeds2015SunkCosts} find no evidence that highly drafted players receive extra time. How evenly a team spreads minutes over a season is associated with its wins \citep{Cardazzi2024Concentration}. Team points have been related to the ice time given to data-driven NHL player types, with a tool that projects reallocations \citep{Chan2012NHLTypes}, and NBA rosters have been chosen by mixed-integer nonlinear programming that values synergies between archetypes \citep{Muniz2023Balanced}. Our coverage program asks a narrower question in each game: how far the retained rotation could restore the absent players' minutes and roles under one budget and each player's capacity. Shot allocation has been evaluated for efficiency \citep{Goldman2011Allocative,Sandholtz2020SpatialAllocation}, and inverse optimisation recovers decision-makers' objectives \citep{Chan2025InverseOptimization,Sandholtz2024RiskPreferences}. Skill sets complement one another \citep{Kuehn2017Complementary,Maymin2013Chemistry}, productivity spillovers matter for team output \citep{Arcidiacono2017Spillovers}, and efficiency is expected to fall as a player's usage rises \citep{Skinner2017Optimal}; coverage holds role rates fixed and so describes a scenario, not what the remaining players would produce.

\section{Data and information timing}

We study the 2023--24 to 2025--26 NBA seasons: 3,690 regular-season games and the three NBA Cup finals, which ESPN lists with them but which count in no standings. The finals contain no case but, like all games, enter player histories and ten-game baselines. The two preceding seasons (2,460 team-games each) form the second sample (Section~7); their reports were parsed with the same parser, adapted to the older layout. They list COVID-19 absences under the league's health and safety protocols, which our reason classes count as medical. Seasons from 2019--20 enter only as player history. The data construction is shared with a separate paper by the same authors on a different question \citep{GuanZhou2026Out}; the two papers share nothing else.

\subsection{Sources}

\begin{itemize}
  \item \textbf{Games and box scores.} ESPN game summaries (8,292 games, 2019--20 to 2025--26): start time, final score and every listed player with minutes and box statistics. The summaries also embed an injury block that is updated after the game; we do not use it.
  \item \textbf{Official injury reports.} Every PDF injury report the league's server holds for the game days of October 2021 to April 2026 from midnight (US Eastern) to 90 minutes before the day's last tip-off (23,485 reports), hourly and, from 22 December 2025, every 15 minutes. Collection stopped at that endpoint, not publication: the server also holds 1,787 later reports of those days, which we did not use. None of them can meet a cutoff six hours or 90 minutes before a game's own tip-off, so they change no selected report and nothing in this paper. Our parser, which assigns each word to a column by its position, parsed every report. On an earlier version of the data (about five reports per game day), its Out list agreed with a parse from our earlier work in 99.8\% of the 5,978 team-games both covered. We time each report by the time printed in its header, the earliest time it can have been public; until 22 December 2025 file names carry only the hour and are 30 minutes earlier (45 minutes for 57 reports of 19 to 22 December 2025), an offset that the examples of \citet{Abramson2021Injury} also show.
  \item \textbf{Transactions and attributes.} The league's player-movement log (6,716 signings, waivers and trades) and season-level listed position and height from Basketball-Reference roster rows, matched to players by team and season (one removed player of the earlier seasons has no roster row), used for position and forecasts.
  \item \textbf{Play-by-play and betting lines.} The ESPN summaries also hold the play-by-play, from which we reconstruct lineup stints (the stretches between substitutions) for the substitution-partner share and a plus-minus measure, and closing spreads for 2021--22, 2022--23 and 2025--26. The stints reproduce the final score exactly in 57.8\% of games (92.3\% within four points for each team); recomputed from the games whose stints pass our validation checks (44\% of the games before 2024--25), both measures give similar estimates and the same conclusions (Supplementary Tables S45 and S47).
\end{itemize}

\subsection{The pregame information set}

The information cutoff is six hours before scheduled tip-off. A player belongs to a team if his most recent event before the cutoff (a box-score listing, a signing, a trade or a claim) ties him to that team and he has not since been waived; a player listed on the team's injury report is added unless he is known to belong to another team. The movement log has no contract expirations, so a player whose last tie to a team is from an earlier season stays on it only if his first record of the new season (a box-score row, a listing, a signing, a trade or a waiver) is a box-score row or listing for that team, or a waiver (a re-signing does not count); this uses records after the cutoff and reconstructs the roster after the fact. A ten-day contract ends after ten days or three team games, whichever is longer, unless renewed, and transactions, dated by day, count from the end of their day (Section~8). Each player's status (Out, Doubtful, Questionable, Probable, Available) and reason come from the last official report whose header time is at or before the cutoff (equal to it in 51\% of cases; Section~8) in which the team's block (its section of the report) was submitted; ruled Out below means this status. Reasons are kept distinct (medical, G League assignment, personal, suspension, rest, trade pending, ineligible, unknown). These classes count ``Coach's Decision'' as medical (25 of 26,554 medical Out rows) and a bare ``Not With Team'' as personal (688 rows); a correction written after the analyses were frozen separates both. It changes no case or removed set and none of the estimates, only one descriptive count in the Supplementary Material, so the tables were not rebuilt with it (Section S4). Player history is each player's last 20 regular-season games (counting the NBA Cup finals) with positive minutes that ended before the cutoff, within 365 days, for any team. Besides this roster rule, two preprocessing steps use information from after the cutoff. Role rates are standardised with the moments of the analysed seasons, so we repeat the allocation test with the previous season's moments (Supplementary Tables S43 and S45). Listed positions and heights are season-level values. Apart from the holdover rule, whose first record of the season can be the case game itself, and the end of ten-day contracts, which can depend on the following days, actual participation, minutes and scores from the same game enter only as outcomes. The rosters and these two steps are thus not as known at the cutoff: the analysis is retrospective, not a simulation of forecasts made at the time.

\subsection{Missing report blocks and the number of cases}

At the six-hour cutoff, 664 team-games (9.0\%) have no submitted block for the team in any report stamped by then. Under the 2025--26 rules most game-day reports are due between 11 a.m.\ and 1 p.m.\ local time \citep{NBA202526InjuryRules}, close to a 1 p.m.\ cutoff for a 7 p.m.\ tip-off. A core player sat out in 50.6\% of team-games with missing blocks and in 47.6\% of the rest, so selective missingness cannot be ruled out. Team-games without a block are left out of case identification rather than treated as healthy. A hybrid sample, added after the first results, fills them with the last report stamped 90 minutes or more before tip-off and serves as a sensitivity analysis.

The 2025--26 season has 1,058 cases (43.0\% of team-games), against 630 (25.6\%) in 2023--24 and 753 (30.6\%) in 2024--25. Three things rose together: more blocks had been filed by the cutoff, core players missed more games according to the box scores, and a core player who sat out was more often already listed Out, for any reason, at the cutoff (Supplementary Table S53).

\subsection{Validation and hand checks}

The authors made two hand checks. First, for 184 team-games of four teams in 2023--24 they reconstructed rosters, report statuses and participation by hand from official transactions and reports, split between the authors. Table~1 compares the automated pipeline with this subset. Of the 89 cases the pipeline finds in these team-games, 74 are cases in the audit. The other 15 fall in team-games whose rosters the audit could not verify; a separate check of their source records confirms 14, and the authors, checking all 15, agree and class the fifteenth, a return-to-competition reconditioning after a non-medical absence, as not medical, although our reason rule counts it; without it the tests barely change (Supplementary Table S3). Second, for 80 team-games from all five seasons, 28 of them from the two earlier seasons, they compared the report the pipeline selected, its Out list, and its statuses and reasons with the source reports. The selected report and the Out list agreed in all 80. Statuses and reasons agreed in 78, and the two disagreements, reasons that ran across a page break, led to a correction of the parser. A further 40 team-games of the two earlier seasons, half of them cases, were then checked in the same way and agreed on every item. A check that needs no hand work: a core player listed Out at the cutoff played in only 8 of 5,171 instances (Supplementary Table S53). Every disagreement, in either sample, is a player listed Out in the report the pipeline selected; the 31 at six hours (28 with no report for the team at the cutoff in the reconstruction) are all in reports stamped exactly at the cutoff and disappear with reports stamped strictly before it (Supplementary Tables S3 and S72).

\begin{table}[H]
\centering\small
\caption{Pipeline versus the hand-reconstructed subset of 184 team-games (six-hour cutoff). The remaining Out disagreements are players the pipeline lists as Out and the audit does not (31 of 734). Ties in reference minutes at the third place (about 1\% of team-games) are broken by player identifier, here as in the pipeline.}
\label{tab:1}
\begin{tabular}{>{\raggedright\arraybackslash}p{0.46\textwidth}ll}
\toprule
Check & Agreement & n \\
\midrule
Rotation players in audited roster found by pipeline & 99.9\% & 2,544 \\
Pipeline rotation players in audited roster & 98.8\% & 2,573 \\
Pregame Out status agreement & 99.1\% & 3,431 \\
Out recall / precision & 100.0\% / 95.8\% & 703 / 734 \\
Identical three-player reference core & 98.9\% & 184 \\
Audited cases recovered & 100\% (74 of 74) & 74 \\
Coverage absolute error, median / 90th percentile & 0.000 / 0.001 & 74 \\
\bottomrule
\end{tabular}
\end{table}

\section{Measures}

\subsection{Reference rotation}

A roster member is supported if his history has at least 5 games and 100 minutes and no more than 20 of the team's scheduled games have passed since his last appearance. For supported player $j$, let $\bar m_j$ be his mean minutes and $u_j = \min(40, q_j)$ his minutes capacity, where $q_j$ is the 90th percentile of his minutes. The reference rotation, which we also call the reference allocation, divides the 240 minutes of a regulation game among the supported players in three steps; the resulting $b_j$ are reference minutes. First, the minutes are shared in proportion to $\bar m_j$. Second, every player whose share exceeds $u_j$ is set to $u_j$ and leaves the pool together with those minutes. Third, the first two steps are repeated on the remaining players and minutes until no share exceeds a capacity (algorithm in Supplementary Section S1). The rotation uses only games that ended before the cutoff and includes every supported player, whether or not he is ruled Out. It is a full-roster baseline, not an estimate of what each player would play with nobody absent. The three players with the largest $b_j$ form the reference core.

\subsection{Role rates}

For each role $k$ (three-point attempts, assists and total rebounds per minute) a player's rate $r_{jk}$ is shrunk, with a weight of 100 pseudo-minutes, toward the pooled rate of his teammates (the player himself excluded):
\begin{equation}
r_{jk} = \frac{n_{jk} + 100\,\bar r_{-j,k}}{M_j + 100}
\end{equation}
where $n_{jk}$ is his count, $M_j$ his minutes in the history window and $\bar r_{-j,k}$ that pooled rate.

\subsection{Role similarity}

Role similarity compares a teammate's profile with that of the absent players. Each rate is standardised over the player rows of all team-games with a core player ruled Out in the seasons analysed, $z_{jk} = (r_{jk} - \mu_k)/\sigma_k$: the three main seasons, the two earlier ones (Section~7), all five in five-season pools, and the two training seasons for the forecasts (Section~5.5; Supplementary Section S1). Let $R$ be the supported players ruled Out at the cutoff for any reason, the removed set: 3.0 per case on average (median 3), the medically absent core players and any others. Its profile is the reference-minute-weighted mean of its members' standardised rates, and similarity is the cosine between the two profiles:
\begin{equation}
\bar z_{Rk} = \frac{\sum_{a\in R} b_a z_{ak}}{\sum_{a\in R} b_a}, \qquad \text{sim}_j = \frac{\sum_k z_{jk}\,\bar z_{Rk}}{\lVert z_j\rVert\,\lVert \bar z_R\rVert} .
\end{equation}

Similarity lies between $-$1 and 1, and its within-case standard deviation (the root mean square deviation from the case mean) is 0.58. It is high when a teammate is above or below average in the same roles as the removed set, whatever his minutes. The removed set's profile is never zero (minimum norm 0.06, median 1.00); without the 29 cases below 0.2 the coefficient of Section~5.1 is almost unchanged (1.10 against 1.09; Supplementary Table S64). An example: when Philadelphia hosted Milwaukee on 25 February 2024, the removed set was the starting centre, Joel Embiid, and one reserve (30.7 reference minutes together). Its profile had few three-point attempts, above-average assists and many rebounds (standardised rates $-$0.71, 0.77 and 1.65). The two reserve big men had the highest similarity (0.76 and 0.65), and one of them, Paul Reed, played 27 minutes against 16.4 reference minutes (Table S55).

\subsection{Role coverage}

The removed set (the supported players ruled Out; Section~4) leaves $L = \sum_{a\in R} b_a$ reference minutes and role demands $D_k = \sum_{a\in R} b_a r_{ak}$. Each retained player $j$ (any other supported player) keeps his reference minutes and can add $x_j \le h_j = \max(u_j - b_j, 0)$, his headroom. Coverage is the largest fraction $\theta$ of the removed minutes and of every role demand that one set of extra minutes can restore at the same time:
\begin{equation}
\begin{aligned}
\text{coverage} = \max_{\theta,\,x}\quad & \theta \\
\text{subject to}\quad & \textstyle\sum_j x_j \le L, \qquad \textstyle\sum_j x_j \ge \theta L, \\
& \textstyle\sum_j r_{jk}\,x_j \ge \theta D_k \quad \text{for every role } k, \\
& 0 \le x_j \le h_j \quad \text{for every retained player } j, \qquad 0 \le \theta \le 1 .
\end{aligned}
\end{equation}

The same minute counts toward every role, so a player cannot cover rebounds and assists with separate minutes. The model describes a scenario and is not an upper bound on real substitutability: it keeps every retained player at least at his reference minutes, lets no one exceed his capacity and holds role rates fixed. In the games themselves, 29.5\% of retained players played fewer minutes than their reference minutes (15.5\% did not play), 15.4\% more than their capacity, and players outside the reference rotation played minutes equal to 16\% of the removed reference minutes (Supplementary Table S46). The shared budget lowers coverage below the smallest separate bound (minutes or one role) by 0.014 on average (median 0; Supplementary Table S46).

For the role-output test we also compute coverage for each role separately, $c_k = \min(1, S_k / D_k)$, where $S_k$ is the largest amount of role $k$ that the retained players can supply with at most $L$ extra minutes within their headroom (Supplementary Section S1). Among the 2,441 primary cases, median joint coverage is 0.951; 39.2\% reach 1.

\section{Empirical design}

\textbf{Cases.} A case is a team-game in which at least one reference-core player is ruled Out for a medical reason at the cutoff. The primary six-hour sample has 2,441 cases in 2,013 games, covering all 30 teams and 348 distinct combinations of team and absent core players. The removed set contains every supported player ruled Out for any reason, on average 3.0 per case; the other supported players are the retained players. Most cases lie inside longer absences: in 59.2\% the same core players were also out in the team's previous game, and in only 3.9\% of cases had every removed player played in that game (Supplementary Table S40). Supplementary Section S1 defines every variable.

\textbf{Planned tests.} Three tests were planned: their specifications, samples, three sensitivity analyses (Section~8) and pass rules were recorded in our version history before any relationship between a role measure and an outcome was computed. That history is our own record; these plans were not registered externally \citep{Nosek2018Preregistration}, unlike the 2026--27 replication (Section~9.2). In earlier work on these data we had seen the final scores of all five seasons, fitted models of player value and team production, and compared scores with betting lines; we had not related role similarity or coverage to any outcome (Supplementary Section S4).

\textbf{Allocation test.} The unit is retained player $j$ in case $i$, and the outcome is his extra minutes, his actual minutes (zero if he did not play) minus his reference minutes:
\begin{equation}
\Delta\text{min}_{ij} = \beta\,\text{sim}_{ij} + \gamma_1 b_{ij} + \gamma_2 h_{ij} + \alpha_i + \varepsilon_{ij} ,
\end{equation}
where $b_{ij}$ and $h_{ij}$ are his reference minutes and headroom in that case and $\alpha_i$ is a case fixed effect. Because $b_{ij}$ is a regressor, $\beta$ is also the coefficient of the same model with actual minutes as the outcome: it compares teammates with equal reference minutes and headroom in one case, not a teammate's minutes with and without the absence, which Section~5.2 compares.

\textbf{Role-output test.} The unit is a case and role, and the outcome is role output: the team's in-game total of the role over its mean in the previous ten games, minus one. The regressors are single-dimension coverage $c_k$, the removed minute share (removed reference minutes divided by 240) and role dummies.

\textbf{Margin test.} The unit is the case and the outcome is the final margin. The regressors are coverage, the removed minute share, lost value, strength difference, a home indicator and rest difference. Lost value weights each removed player's reference minutes by his box-score efficiency per minute in his history games; absence value is the same quantity for every team-game with a report block. Strength difference is the difference between the two teams' mean margins over their previous ten games, and rest difference the difference in days since each team's previous game, each capped at three (Supplementary Section S1).

\textbf{Pass rule and inference.} Each test passes if its key coefficient is positive with two-sided $p<$ 0.05; it is called robust to measurement if it also passes in at least 8 of 10 alternative definitions, a rule written after the first results (Section~8). Standard errors (SE) are clustered by team (30 clusters, CR1 correction, t-tests with 29 degrees of freedom; fewer clusters in some pair comparisons). Adjusted for the three planned tests (Holm), the role-output $p$ is 0.14; the exploratory analyses are not corrected for multiplicity. Because 30 clusters are few \citep{MacKinnon2023Guide}, we also report wild cluster restricted bootstrap p-values \citep{Cameron2008Bootstrap} with 9,999 draws and Webb weights \citep{Webb2023Reworking}; some later checks use Rademacher weights. With absorbed fixed effects, all nested within teams, the small-sample factor counts only the explicit regressors (\citealp{Correia2015Singletons}; see \citealp{CameronMiller2015Guide}); counting the absorbed effects too would widen the standard errors we report by 4\% to 13\%, or 18\% to 60\% with teammate or removed-set effects (Supplementary Table S58).

\textbf{Exploratory analyses.} Everything else is exploratory; a few later analyses had a pass rule written before they were run, and most were added after the main results were known. Supplementary Section S4 and Table S71 give the planned tests, their verdicts and the deviations that bear on them; the analysis record released with the code gives each plan, its timing and every deviation. A replication of the allocation test, without and with listed position and height, on the 2026--27 season was registered on the Open Science Framework on 30 September 2026, before the season began, and will be judged once, at the end of the regular season. A final exploratory analysis (Sections~5.3 and~6.3), decided after all other results were known, was specified before it was run in a working log, not in the version history (Supplementary Section S4).

\section{Results: role similarity and minutes}

Table~\ref{tab:planned} gives the planned tests and their verdicts; this section and the next discuss the allocation test and the coverage tests in turn, and Section~7 the earlier seasons.

\subsection{Minutes tilt toward similar teammates}

Within a case, teammates whose role profile is closer to that of the absent players play more minutes than teammates with the same reference minutes and headroom. In the primary sample, a one-unit increase in role similarity is associated with 1.09 more minutes (SE 0.18; $p<$ 0.0001; wild bootstrap $p \le$ 0.0001; 28,118 player-cases), holding reference minutes and headroom fixed. One within-case standard deviation of similarity (0.58) thus corresponds to about 0.63 minutes more than such a teammate receives. Figure~1 shows the relationship net of the case, reference minutes and headroom. With splines, deciles or decile-by-quintile cells for reference minutes and headroom, the coefficient changes by at most 5\% (1.14; earlier seasons 8\%, 1.29; exploratory; Supplementary Table S65).

\begin{table}[H]
\centering\footnotesize
\caption{The planned tests, six-hour cutoff: coefficient and 95\% interval (team-clustered; for the margin test, coverage per 0.1 with the 90\% interval of the equivalence test), in the main seasons (2023--24 to 2025--26, 2,441 cases), the earlier seasons (2021--22, 2022--23, 1,466 cases) and all five seasons (3,907 cases). The verdict applies the pass rule of Section~4 to the main seasons; the position-and-height row (P3 addendum A) was specified after the first results of the other rows and before its own. Details in Supplementary Tables S5 to S8, S21, S23, S30 and S71.}
\label{tab:planned}
\begin{tabular}{@{}>{\raggedright\arraybackslash}p{0.22\textwidth}lll>{\raggedright\arraybackslash}p{0.12\textwidth}@{}}
\toprule
Test & Main seasons & Earlier seasons & Five seasons & Verdict (main) \\
\midrule
Allocation: role similarity and extra minutes & 1.09 [0.72, 1.46] & 1.20 [0.67, 1.72] & 1.14 [0.85, 1.43] & passes\newline ($p<$ 0.0001) \\
Allocation, with position and height & 0.89 [0.51, 1.28] & 0.91 [0.40, 1.41] & 0.92 [0.62, 1.22] & passes\newline ($p<$ 0.0001) \\
Role output: single-dimension coverage & 0.054 [$-$0.005, 0.114] & 0.100 [0.047, 0.153] & 0.070 [0.029, 0.112] & does not pass\newline ($p=$ 0.070) \\
Role output, with the benchmark (exploratory) & $-$0.053 [$-$0.126, 0.021] & 0.068 [$-$0.009, 0.144] & $-$0.002 [$-$0.064, 0.060] & -- \\
Margin: coverage per 0.1, 90\% interval & $-$0.10 [$-$0.58, +0.39] & +0.24 [$-$0.31, +0.80] & +0.05 [$-$0.32, +0.42] & does not pass\newline ($p=$ 0.73) \\
\bottomrule
\end{tabular}
\end{table}

Because reference minutes enter both the outcome and the regressors (Section~4), this is a conditional difference between teammates of one game: it says who plays longer in that game, not who gains minutes because of the absence. The comparisons of Section~5.2 come closer to the second question: 1.54 minutes per unit of similarity between a teammate's games without and with a given core player, but 0.77 (95\% interval $-$0.09 to 1.62) in the 168 spells whose windows remove a linear drift.

\begin{figure}[H]
\centering
\includegraphics[width=0.6\textwidth]{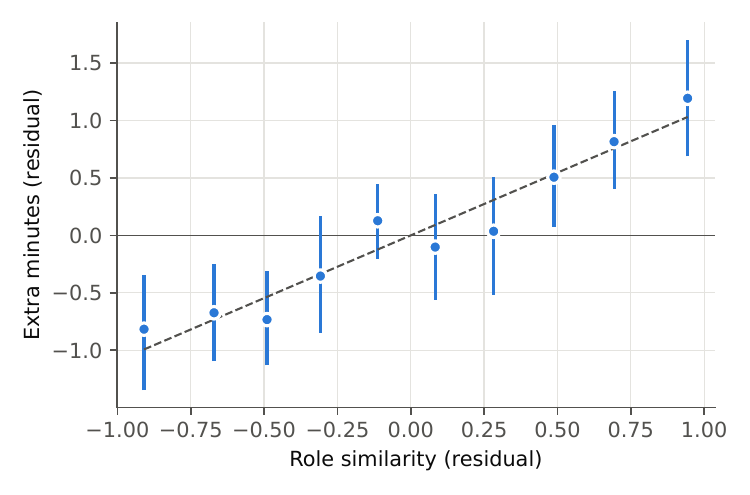}
\caption{Partial relation between role similarity and extra minutes, net of case fixed effects, reference minutes and headroom (primary sample, 28,118 player-cases). Points are means within deciles of residual similarity; bars are 95\% team-cluster bootstrap intervals; the dashed line is the regression slope (1.09 minutes per unit).}
\label{fig:similarity}
\end{figure}

The association is small. The partial R-squared of similarity is 0.45\%, against 3.4\% for reference minutes and headroom together, and extra minutes have a within-case standard deviation of 9.6 minutes (Supplementary Table S52). The pattern describes allocation and is not evidence of how coaches decide. Because the profiles come from recent games that often include the current absence (Section~4), similarity describes current roles, possibly already adapted to the absence. Two checks point the other way: Section~5.3 also measures similarity before the absence (1.01 in the pair comparison without position terms), and the gradient is larger in the first game of a run of absences than in the games that follow (Section~5.4). Split into parts (exploratory; Supplementary Table S64), it reflects both whether a teammate plays (1.1 percentage points per standard deviation; 85\% play) and his minutes when he does (0.50 minutes per standard deviation). The coefficient barely moves when the teammate's own pregame status is added (1.00) or teammates listed Questionable or Doubtful are left out (1.08; exploratory; Supplementary Table S61).

\subsection{Games without and with the absent player}

Extra minutes are measured against reference minutes, and reference minutes come from a history in which the absent players are often already absent (Section~4). The coefficient above is therefore a deviation from a mixed baseline, not the difference between a teammate's minutes without and with a given player. An exploratory analysis makes that comparison directly, with an outcome that does not use the reference allocation (Supplementary Table S49). The unit is a pair. One member is a core player who, within a team-season, was ruled Out for a medical reason in at least five cases and played in at least five of the team's other games with a report block. The other is a teammate observed in at least three games of each kind. The outcome is the teammate's mean minutes in the games without the core player minus his mean minutes in the games with him, where a game in which the teammate did not play counts as zero and a game for which he was himself ruled Out is left out. We regress this difference on the similarity of the two players (the cosine of their standardised rates, averaged over the games without the core player), with a fixed effect for each absent player in each team-season. The teammate's mean reference minutes and headroom in those games are controls; the reference rotation still defines who is a core player and which teammates enter.

The primary sample gives 2,653 such pairs, built around 199 combinations of absent player, team and season. Teammates play 2.9 minutes more on average in the games without the absent player, and 1.54 minutes more per unit of similarity to him (SE 0.19; 95\% interval 1.15 to 1.93; wild bootstrap $p \le$ 0.0001). The estimate is 1.30 (SE 0.26) when the absent player is the only core player out and no core player is out in the comparison games, and 1.41 (SE 0.29) in the two earlier seasons.

Without a teammate effect, a teammate who gains minutes whenever any core player is absent would contribute to this estimate. With a fixed effect for each teammate in each team-season, each teammate is compared only with himself across the absent players he is paired with, in the 2,168 pairs in which the teammate (828 teammate-seasons) is paired with two or more absent players. The estimate is 1.23 (SE 0.25, or 0.33 with the absorbed effects counted; 95\% interval 0.73 to 1.73; wild bootstrap $p=$ 0.0003), against 1.52 on the same pairs without teammate effects, and 1.62 (SE 0.42) in the earlier seasons (Supplementary Table S56).

The comparison is still observational: games without and with a player differ in date and opponent, and a teammate's place in the rotation can change over a season. Windows of three games on each side of the start and of the end of a spell of absence cancel a drift linear in the team's game number, though not necessarily one linear in calendar time (Supplementary Table S60). In the 168 spells with both windows the gradient is 0.77 (SE 0.42; $-$0.09 to 1.62), too imprecise to tell. In the games before a spell, which involve no change of status, no gradient is detected; in the earlier seasons the windows give 2.43 (1.72 to 3.15).

\subsection{Position, height, past substitution and player effects}

Similarity might simply recode listed position or the absent players' usual backups. Two kinds of evidence bear on this: how the coefficient changes when position controls are added within one sample, and whether what remains is the same across designs (Table~3; Supplementary Section S6 gives the full account). Three position groups (guard, forward, centre) and the height gap leave 82\% of the coefficient (Supplementary Table S6). On one common sample (Table~3, rows 1 to 7), adding a substitution-partner share, how often the teammate was on court during an absent player's bench possessions in his recent games, leaves 72\%; indicators for every pairing of the teammate's five listed positions with the removed set's main position (the one holding most of its reference minutes) leave 63\%; and the share of the removed reference minutes at each listed position, interacted with the teammate's position, leaves 45\%, or 53\% with player effects.

Two designs describe listed position completely, although listed position is not actual role. In the 455 cases whose removed players share one listed position, the pairings and the height gap take the coefficient from 1.05 to 0.37 ($-$0.44 to 1.18), and adding the partner share to 0.20 (row 9). In the pair comparison of Section~5.2, on the 1,278 pairs with a game before the absence (1.30 there without position terms, against 1.23 on all pairs), with the 25 pairings of the two players' listed positions, their height gap, absent-player and teammate effects and similarity measured before the absence, it falls from 1.01 to $-$0.02 ($-$0.72 to 0.67; rows 10 and 11). No association beyond position is detected in these designs, but one of 0.54 minutes per unit of similarity, the size the plan of the pair analysis called meaningful (0.3 minutes per standard deviation, about half the 0.63 of Section~5.1), cannot be excluded (one-position minus other cases: $-$0.56, $p=$ 0.22); with absent-player effects only the pair estimate is 0.60 ($-$0.05 to 1.25), or 0.72 (0.19 to 1.25) on all 1,788 pairs with similarity before the absence, and in the earlier seasons the pair model gives 0.85 (row 13). The cases and the pairs differ in more than the position control (unit, sample, outcome, timing of similarity, fixed effects), so the gap between them is not the effect of a more complete control alone.

\begin{table}[H]
\centering\small
\caption{Role similarity with position, height, substitution and player controls, in three designs that differ in more than the controls. Panel A, rows 1 to 7: the allocation test on one common sample (27,896 player-cases in 2,423 cases with every teammate control available); player effects are player-by-team-by-season fixed effects; share kept is relative to the similarity-only model with the same fixed effects. Panel B, rows 8 and 9: the 455 of these cases (5,775 player-cases) in which every removed player has the same listed position. Panel C, rows 10 to 13: the comparison of games without and with the absent player (Section~5.2). All rows are exploratory. Other models and the earlier seasons are in Supplementary Tables S45, S50, S56, S57, S58, S59 and S63.}
\label{tab:2}
\begin{tabular}{@{}>{\raggedright\arraybackslash}p{0.55\textwidth}lll@{}}
\toprule
Model (all with reference minutes and headroom) & Similarity (SE) & 95\% interval & Share kept \\
\midrule
\textbf{A. All cases (2,423), case effects} &  &  &  \\
(1) Similarity only & 1.10 (0.18) & 0.73 to 1.47 &  \\
(2) Three position groups, height gap, partner share & 0.79 (0.19) & 0.41 to 1.17 & 72\% \\
(3) Every position pairing, height gap, partner share & 0.69 (0.17) & 0.35 to 1.04 & 63\% \\
(4) Positions of all removed players, height gap, partner share & 0.49 (0.17) & 0.15 to 0.84 & 45\% \\
\textbf{A, continued: case and player effects} &  &  &  \\
(5) Similarity only & 1.40 (0.20) & 0.98 to 1.82 &  \\
(6) Every position pairing, height gap, partner share & 0.93 (0.23) & 0.47 to 1.39 & 66\% \\
(7) Positions of all removed players, height gap, partner share & 0.75 (0.22) & 0.30 to 1.19 & 53\% \\
\textbf{B. One-position cases (455), case effects} &  &  &  \\
(8) Similarity only & 1.05 (0.32) & 0.40 to 1.70 &  \\
(9) Every position pairing, height gap, partner share & 0.20 (0.41) & $-$0.64 to 1.04 &  \\
\textbf{C. Pairs (1,278; row 13: 914), similarity before the absence} &  &  &  \\
(10) Absent-player and teammate effects & 1.01 (0.32) & 0.36 to 1.66 &  \\
(11) Both effects, position pairings and height gap & $-$0.02 (0.34) & $-$0.72 to 0.67 &  \\
(12) Absent-player effects only, position pairings and height gap & 0.60 (0.32) & $-$0.05 to 1.25 &  \\
(13) Earlier seasons (914), as row 11 & 0.85 (0.58) & $-$0.34 to 2.04 &  \\
\bottomrule
\end{tabular}
\end{table}

Case fixed effects absorb who is absent, but not a teammate's place in the rotation. With player-by-team-by-season fixed effects the coefficient rests on changes in the same player's similarity across his cases, almost all between cases with different removed sets: it is 1.38 (0.95 to 1.81) on all primary cases and 0.93 with every position pairing, the height gap and the partner share (row 6), so a place in the rotation that is the same in all of a player's cases in a season does not account for the association; player effects do not absorb his being the regular replacement of one particular teammate, which the partner share addresses only in part.

Two exploratory analyses ask which roles carry the gradient and what happens when position and role disagree (Supplementary Tables S66 to S68). Entered jointly as separate regressors, with the teammate's own scaled rates as controls, the three role components of similarity give 0.75 minutes per unit for three-point attempts, 1.01 for assists and 1.76 for rebounds, and their equality is not rejected ($p=$ 0.070); with the controls of Table~3, row 3, and own status the rebound component remains positive (1.03; 0.52 to 1.55), and it and the assist component come from absences of players who rebound much and pass little, mostly forwards and centres. In the 280 cases with one removed player that contain both a teammate of his position group with low similarity and one of another group with high similarity, the second plays 0.15 minutes more ($-$0.83 to 1.13), an interval too wide to tell which criterion the allocation follows.

\subsection{Placebo and subsamples}

Figure~2 collects the main estimates of this section and of the earlier seasons (Section~7). Every real-absence row without the position pairings of Section~5.3 has a positive coefficient whose 95\% interval excludes zero, and both placebo intervals include zero.

\begin{figure}[htbp]
\centering
\includegraphics[width=0.9\textwidth]{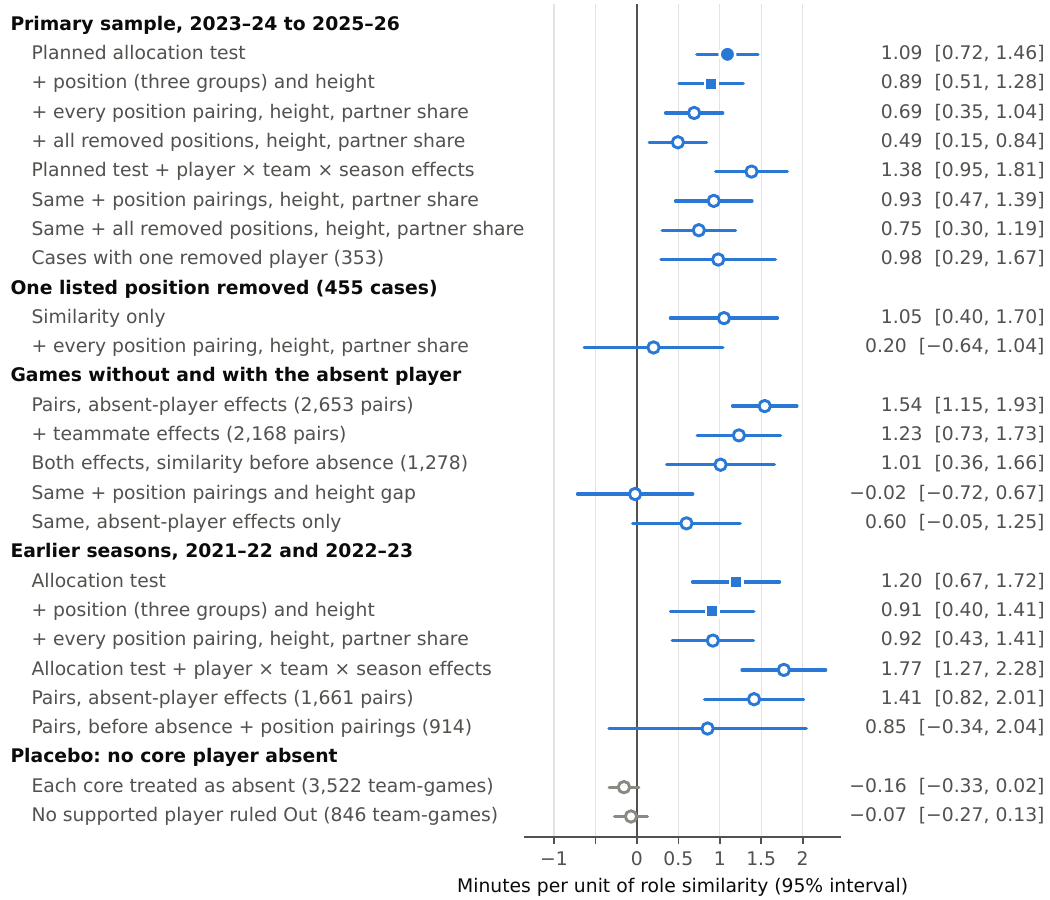}
\caption{The similarity coefficient across specifications and samples: minutes per unit of role similarity, with 95\% intervals (coefficient $\pm$ 2.045 team-clustered standard errors; for the pair comparisons with fewer than 30 teams, the $t(G-1)$ quantile for their number of teams $G$). Filled circle: the planned allocation test; filled squares: tests whose pass rule was written before they were run; open circles: exploratory checks. Pair rows: outcome a difference of mean minutes, similarity to that player alone (Section~5.2); other rows: extra minutes, similarity to the whole removed set. Rows use different samples; Table~3 provides same-sample comparisons. Sources: Sections~5 and 7 and Supplementary Tables S6, S18, S23, S48 to S50, S56, S57, S59 and S63.}
\label{fig:summary}
\end{figure}

In a placebo with the 3,522 team-games in which all three core players took part, treating each in turn as absent, the similarity gradient is slightly negative ($-$0.16 per unit, SE 0.09; $-$0.21, $p=$ 0.026, with one pseudo-absent core player drawn at random), as expected if similar teammates compete for the same minutes when the core player plays or if reference minutes already reflect earlier absences, and the real-minus-placebo difference is 1.25 (SE 0.20, $p<$ 0.0001; Supplementary Table S18). Because placebo similarity refers to one player, a closer contrast takes the 353 cases with one removed player (0.98, $p=$ 0.007) and the 846 placebo team-games in which no supported player was ruled Out ($-$0.07, SE 0.10): the difference is 1.05 (SE 0.35, $p=$ 0.005; wild bootstrap $p=$ 0.007; exploratory; Supplementary Table S48).

The association also holds with Euclidean distance and in the 406 cases in which only core players were removed, and it is larger in the first game of a run of absences (1.32) than in the games that follow, where it remains positive (Supplementary Tables S35, S40 and S54).

\subsection{Forecasts}

In the retrospective allocation models we tested, similarity did not yield a statistically detectable reduction in the mean absolute error of minute forecasts. Regressions fitted to the hybrid cases of 2023--24 and 2024--25, with role rates standardised on those two seasons alone, and evaluated on the 1,139 of 2025--26 (a season whose outcomes we had seen, though not used in fitting) had mean absolute errors of 7.175 minutes with similarity, 7.166 with listed position (three groups) and height, and 7.201 for allocation in proportion to reference minutes, with no difference distinguishable from zero, and adding similarity to the model with the five listed positions changes the error by $-$0.004 minutes (95\% interval $-$0.017 to 0.011; Supplementary Table S51); an inverse-optimisation model did not meet the pass rule set for it in advance (7.178; Supplementary Tables S14 and S71). None can place a retained player below his reference minutes, although 29.5\% played less (Section~3.4). The comparison is thus limited to models of this kind, fitted and evaluated on the hybrid sample with rosters that use later records (Section~2.2): it shows no advantage for similarity, not that similarity has no predictive value.

\section{Results: role coverage}

\subsection{Team role output: coverage adds nothing detectable beyond the role share}

Single-dimension coverage varies little: it equals its cap of 1 in 77\% of case-roles, and at the estimated slope moving from its 10th percentile (0.85) to 1 corresponds to role output higher by 0.8\% of the ten-game baseline, against an outcome standard deviation of 18\%. Whatever its sign, the association is therefore small. The role-output test does not meet its threshold (Table~2): a one-unit increase in coverage goes with a team role total 5.4\% higher relative to its baseline (SE 2.9\%; $p=$ 0.070; wild bootstrap $p=$ 0.071). It is significant in 2 of the 10 measurement definitions, short of the 8 of 10 that the plan's robustness rule requires (Supplementary Table S71); of its three planned sensitivity samples one passes (core players ruled Out for any reason: 0.058, $p=$ 0.045; Supplementary Table S8).

\begin{table}[H]
\centering\small
\caption{Exploratory versions of the role-output test, main seasons, six-hour cutoff: coefficient of $c_k$ ($p$). The planned test and its five-season and earlier-season versions are in Table~2; the benchmark is defined in the text and in Supplementary Section S1. Supplementary Tables S30, S34 and S43.}
\label{tab:3}
\begin{tabular}{ll}
\toprule
Specification & Coefficient of $c_k$ ($p$) \\
\midrule
With the benchmark (95\% interval $-$0.126 to 0.021) & $-$0.053 (0.15) \\
With the absent players' share of the role instead & $-$0.048 (0.15) \\
With the benchmark capped at each player's headroom & $-$0.044 (0.24) \\
Healthy baseline, with the benchmark & $-$0.074 (0.071) \\
First game of each episode, with the benchmark & $-$0.042 (0.41) \\
Placebo (all core players took part), planned inputs & 0.033 (0.021) \\
\bottomrule
\end{tabular}
\end{table}

When the absent players carried much of a role, $c_k$ is low, but their teammates would also supply less of that role if the freed minutes were simply shared out. As a benchmark we computed, for each case and role, the proportional change in role output if the removed minutes went to the retained players in proportion to their reference minutes, each at his historical rate. Given the removed minute share, the benchmark is almost a linear function of the absent players' share of the role (partial correlation $-$0.98), so we read it as that share. It predicts role output (0.21 per unit, $p<$ 0.0001), and with it as a control the coefficient of $c_k$ is $-$0.053 ($p=$ 0.15); the same holds with the role share itself, with the benchmark capped at each player's headroom, and over five seasons (Tables~2 and~4; Supplementary Table S30). The ten-game baseline is itself partly adapted: in 98.5\% of cases it includes a game without a removed player (Supplementary Table S40), which pulls role output toward zero in continuing absences; with a healthy baseline the coefficient with the benchmark is $-$0.074 ($p=$ 0.071), and in the first game of each episode $-$0.042 (Supplementary Tables S30 and S34). A placebo in the 3,522 team-games in which all three core players took part gives a same-signed slope with the planned inputs (0.033, $p=$ 0.021), and the difference between real cases and placebo is not significantly positive in any version we compared; in one it is significantly negative (Supplementary Tables S43 and S44).

\subsection{The final margin: no consistent association}

Absences coincide with lower margins: team-games with a core player ruled Out for a medical reason have a margin 2.26 points lower than the same team's other games in that season. The margin test asks whether coverage adds to the value of the players who are missing.

In the primary sample, conditional on the removed minute share, lost value, strength, home and rest, the coverage coefficient is $-$0.10 points per 0.1 of coverage with a 95\% interval of [$-$0.68, +0.49]: the interval still includes a negative association of half a point per 0.1, and the 90\% interval does not fall within the equivalence bounds of $\pm$0.5 (see below). Per unit of coverage this is $-$0.99 points (SE 2.86; $p=$ 0.73; wild bootstrap $p=$ 0.74). In the hybrid sample it is $-$0.01 [$-$0.53, +0.51]. No sample, baseline or measurement variant of the planned model yields $p<$ 0.05 for coverage (Supplementary Figure S1); of the exploratory versions we tried, a few do, none of them planned: a five-season hybrid version with a lineup-based value measure ($p=$ 0.043; wild bootstrap 0.055), the 37 cases whose removed players had played each of the team's last ten games ($p=$ 0.039) and the 2021--22 slope of a model with season-specific slopes ($p=$ 0.045; Supplementary Tables S40 and S33). The partial relationship over all five seasons has no clear gradient (Supplementary Figure S2).

How precise is this null? At its standard error, the test had 80\% power only for effects of at least 0.83 points per 0.1 of coverage, about 1.4 points across the interquartile range of coverage (0.17). Two one-sided tests \citep{Schuirmann1987,Lakens2017Equivalence} against $\pm$0.5 points per 0.1, a bound set after the main estimate was known, do not establish equivalence in the primary sample (90\% interval [$-$0.58, +0.39] per 0.1); over five seasons they do so in the planned specification and in some variants (the hybrid sample; RAPM beside the box-score value), but not with the opponent's absence value controlled (Section~7; Supplementary Tables S21 and S26).

Further checks leave this unchanged: other controls, fixed effects and measures of absence value ($p=$ 0.06 to 0.93 in the six-hour models of Supplementary Tables S12, S20, S26, S35 and S42), and other treatments of teammates listed Questionable or Doubtful (exploratory; Supplementary Table S62). A follow-up model with a pass rule written in advance passed (in the hybrid sample the value the retained rotation could supply predicts the margin, $p=$ 0.007), but that measure correlates 0.98 with removed minutes and adds nothing against the closing spread in 2025--26 (Supplementary Tables S15 and S16).

\subsection{Coverage when teammates' minutes may be cut}

Coverage keeps every retained player at or above his reference minutes (Section~3.4), although 29.5\% of retained players played fewer. In an exploratory extension (Section~4; Supplementary Section S6 and Tables S69 and S70) we let retained players' minutes fall below their reference minutes within a budget of total cuts and ask what share of the absent players' minutes and roles can then be restored at once. Given the retained players, their capacities and fixed role rates, 39.2\% of cases are fully covered without cuts, 23.4\% once cuts are allowed (median smallest total cut 9.3 minutes) and 37.4\% not even with unrestricted cuts; allowing cuts closes 43\% of the summed shortfall, 27\% to 60\% under other cut floors, capacity rules and rate shrinkage. Whether the required cut tracks actual rotations is not established: of two planned comparisons one is positive and the other imprecise and based on a baseline that is not comparable in structure, so the criterion fixed in advance (both positive and the baseline comparison's interval excluding zero) is not met.

\section{The same tests in two earlier seasons}

We built the seasons 2021--22 and 2022--23, in which coverage and role similarity had never been related to any outcome, with the same code and ran the planned tests (Table~2) under a rule written down in advance: a test replicates if its coefficient has the original sign and $p<$ 0.05. This is a second sample from the same pipeline and the same authors, not an independent replication.

The allocation test replicates with a similar point estimate in the two seasons pooled, and most of it again survives position (three groups) and height; it is significant in each of the two seasons alone (2022--23: 1.06, $p=$ 0.001; Supplementary Table S38). The exploratory checks of Section~5 agree (1.41 and 1.62 in the without-and-with comparison, 1.77 with player-by-team-by-season effects), and the position results are mixed as in the main seasons: 0.92 (0.43 to 1.41) with every pairing on cases, 0.85 ($-$0.34 to 2.04) in the pair comparison (Section~5.3; Supplementary Tables S48 to S50, S56, S57 and S63).

In the earlier seasons the role-output test meets its planned threshold, also without the 177 six-hour cases that involve a COVID-19 protocol absence (0.089, $p=$ 0.009; the allocation test 1.13; Supplementary Table S33), but with the benchmark the estimate is 0.068 and its interval includes zero. The margin test remains null: over five seasons the planned specification meets the $\pm$0.5 bound of Section~6.2 (Table~2), and with the opponent's absence value controlled it does not (+0.18; $-$0.24 to +0.59).

\section{Robustness and data quality}

\textbf{Measurement definitions.} We reran every step for ten alternative definitions on the hybrid sample (Supplementary Table S10 and Figure S1). The similarity coefficient ranges from 0.81 to 1.12 and is significant in all ten; the coefficient of $c_k$ ranges from 0.028 to 0.064 and is significant in two. The three sensitivity analyses fixed with the plan (without previous-day reports, which removes no case; core players ruled Out for any reason; without full coverage) leave the allocation test unchanged or larger (1.09 to 1.35) and give role output 0.049 to 0.058 ($p=$ 0.045 to 0.093; Supplementary Tables S5 and S8).

\textbf{Inference and seasons.} Wild bootstrap p-values are within 0.03 of the clustered t-test p-values for every planned test (Supplementary Table S9), and clustering by team and game \citep{Cameron2011Multiway} changes no verdict, while clustering by game alone gives role output $p=$ 0.046 (Supplementary Table S32). The allocation test is significant in each main season, with a falling coefficient (1.56, 1.13 and 0.80), but the linear trend is not ($-$0.36 per season, $p=$ 0.060), nor over five seasons ($-$0.11, $p=$ 0.31; Supplementary Table S38).

\textbf{Data corrections.} After the planned tests had been run and the 2026--27 test registered, an audit of the code and a later review found three defects in the rosters of that build: ended contracts, and unrenewed ten-day contracts, were not ended (a player who signed nowhere else kept reference minutes on his old team for about 20 games of the next season; 380 of that build's 2,440 cases), and transactions were dated at noon UTC, before most cutoffs. We specified the corrections (Section~2.2), reran every analysis and, with both sets of results known, made the corrected data the paper's data, although this took the role-output test below its threshold ($p=$ 0.038 on the earlier build); every result here and in the supplement uses them unless a supplementary table compares data versions. Seven other significance judgements differ between the two builds, and on the final data, unlike on the earlier build, clustering by game alone changes the verdict of the role-output test (Supplementary Table S72). In 51\% of cases the selected report's header time, a lower bound on publication, equals the cutoff; with reports strictly before it the tests give 1.01, 0.052 ($p=$ 0.085) and $-$0.08 per 0.1.

\section{Discussion and conclusions}

\subsection{What the evidence shows}

Minutes tilt modestly toward teammates whose statistical role resembles that of the absent players, in the main and the earlier seasons; the placebo gradient is slightly negative.

What the evidence does not settle is whether role similarity carries information beyond listed position. Over all cases 45\% to 72\% of the coefficient remains after position, height-gap and partner-share controls and is distinguishable from zero (Section~5.3). In the one-position cases and the pair model with both players' effects, no additional association is detected in the main or the earlier seasons (Sections~5.3 and 7; Supplementary Table S56), but the intervals are wide, all 1,788 pairs with absent-player effects only give 0.72 (0.19 to 1.25), and these designs differ from the case model in more than the control. The evidence supports a small, repeatable gradient, not that role describes the allocation better than listed position.

Coverage adds nothing detectable to team role output beyond the absent players' share of each role, and nothing consistent to the margin (Section~6). The measure is generous (39.2\% of cases at full coverage) and correlated with the removed minute share and absence value; other dimensions or lineups may matter. Splitting similarity by role and letting coverage cut minutes (Sections~5.3 and~6.3) do not change these readings: the role components do not differ clearly once the teammate's own rates are controlled; the rebound component stays positive with the position controls of Table~3, row 3, but comes from frontcourt absences, so these exploratory analyses do not settle whether role adds to listed position; and how much of the coverage shortfall cuts close depends on capacity and cut-floor assumptions.

\subsection{Limitations}

\begin{itemize}
  \item \textbf{Not causal.} Absences are not random: core players listed Questionable or Doubtful sat somewhat more often when the proportional benchmark said more of their role could be absorbed ($p=$ 0.04; Supplementary Table S42), and neither player nor teammate effects remove a change of place in the rotation that coincides with who is absent (Sections~5.2 and 5.3). Profiles may already reflect the absence (Section~5.1).
  \item \textbf{Seen outcomes.} Every result comes from seasons whose outcomes we had seen. The registered 2026--27 allocation test (amended before its first game to make the corrected code of Section~8 primary) has, at a true effect of 0.78 minutes per unit (its smallest effect of interest), power 0.48, 0.81 or 0.99 if that season is as precise as 2023--24, 2024--25 or 2025--26 (SE 0.40, 0.27, 0.18), and the effects it detects with 80\% power at those precisions are 1.15, 0.77 and 0.52 (Supplementary Table S43); its position rule, three groups and height, cannot settle whether similarity adds to listed position.
  \item \textbf{Sample.} Only supported players (Section~3.1) enter, so long-term absentees, newcomers and players outside the rotation are partly missing.
  \item \textbf{Measurement.} Three box-score dimensions miss usage, defensive assignments, ball handling and spacing. A report's header time only bounds when it became public (Section~8). Coverage, also with cuts, holds role rates fixed.
  \item \textbf{Post-hoc checks.} Only three tests were planned; most checks were specified after related results had been seen. With so many models, isolated p-values near 0.05 should be read accordingly \citep{GelmanLoken2014}.
\end{itemize}

\subsection{What it means}

The paper makes two contributions. First, in pregame absences it measures a small gradient of minutes along a continuous role-similarity measure, which recurs in a second sample, and asks whether it adds to listed position and past substitution. Same-position gains after absences are known \citep{Nishihata2026ValueChanceV2}; new here are the pregame-report timing and the continuous role measure, which extend the question of how an absence shifts co-workers' working time \citep{HoeyPeetersVanOurs2023}. Whether role adds to listed position depends on how position is controlled and on the design (Section~5.3): an analysis that stopped at coarse controls would claim more than these data support. Second, \citet{GlazerBinneySeth2025Weighted} note that their NFL injury-burden measure treats every backup as replacement level and leave team-specific replacement level to future work; coverage is one such attempt, built from box scores, and in the models we tested it shows no stable incremental association, an answer for this measure, not for the question.

\section*{Data and code availability}

Code, tests, analysis plans and a reproduction guide will be made public upon publication. The sources' terms of use restrict redistribution, so we will release derived tables, with a script that reproduces Tables 2 and 3 from them, the main seasons' game identifiers and the SHA-256 hash of every injury report. The 2026--27 replication is registered at \url{https://osf.io/kzfyu} (under embargo).

\section*{Declarations}

\textbf{Funding.} This research received no specific grant from any funding agency in the public, commercial or not-for-profit sectors.

\textbf{Competing interests.} The authors declare no competing interests.

\textbf{Author contributions (CRediT).} Tianyu Guan: Conceptualization; Methodology; Software; Validation; Formal analysis; Writing -- original draft; Writing -- review \& editing; Project administration. Yang Zhou: Conceptualization; Methodology; Software; Validation; Data curation; Writing -- review \& editing. Both authors read and approved the final manuscript.

\textbf{Supplementary Material.} The Supplementary Material follows the references.

\setlength{\bibsep}{0pt}
\bibliographystyle{apalike}
\bibliography{refs}

\clearpage
\includepdf[pages=-]{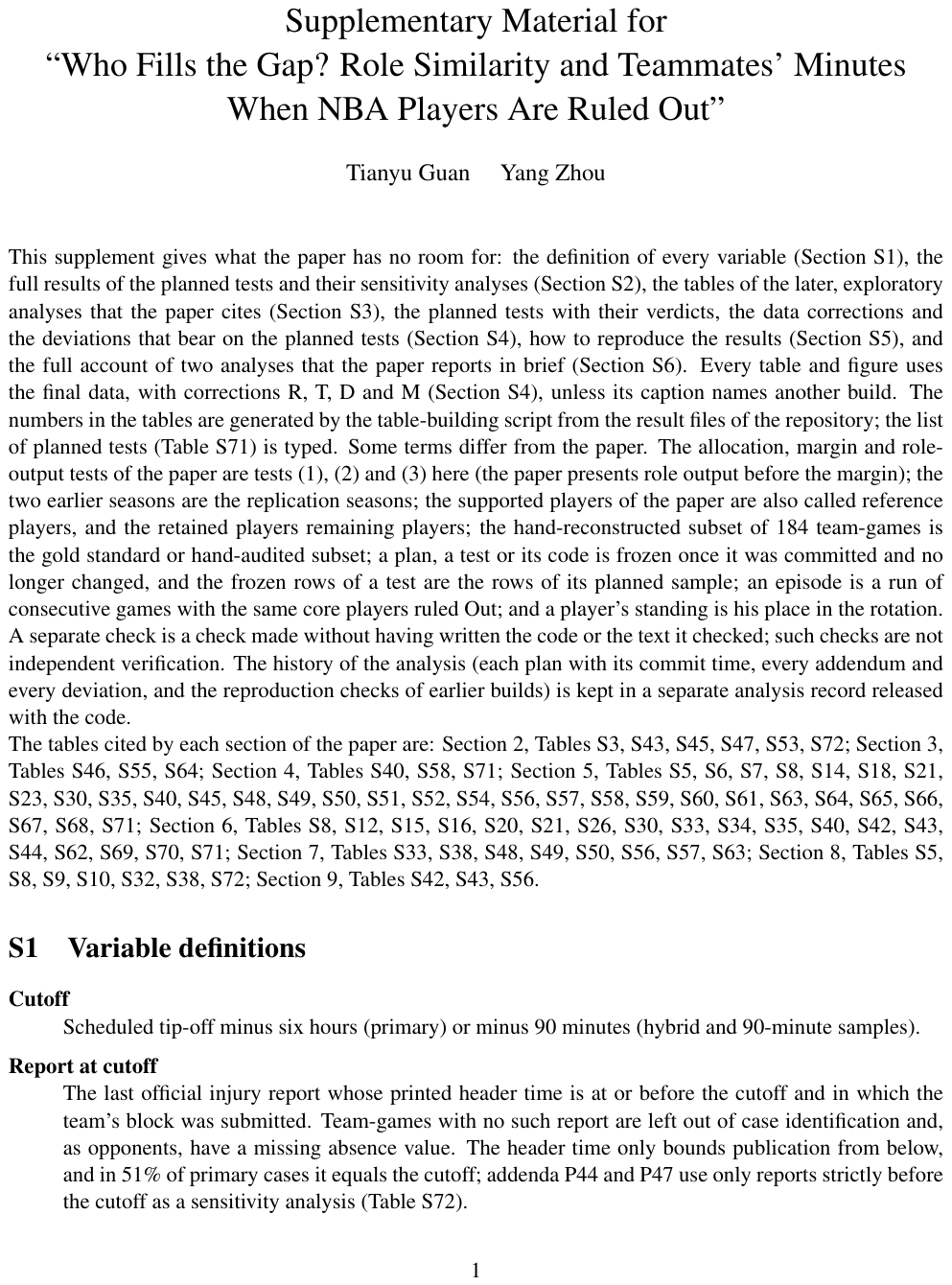}
\end{document}